# Why the Creative Process is Not Darwinian

Commentary on
**'The Creative Process in Picasso's *Guernica* Sketches: Monotonic Improvements versus Nonmonotonic Variants' by Dean Keith Simonton**

LIANE GABORA

UNIVERSITY OF BRITISH COLUMBIA

Address for Correspondence:
University of British Columbia
Okanagan campus, 3333 University Way
Kelowna BC, V1V 1V7, CANADA
email: liane.gabora[at]ubc.ca
Phone: (250) 807-9849
Fax: (250) 470-6001

ABSTRACT

Simonton (2006) makes the unwarranted assumption that nonmonotonicity supports a Darwinian view of creativity. Darwin's theory of natural selection was motivated by a paradox that has no equivalent in creative thought: the paradox of how change accumulates when acquired traits are not inherited. To describe a process of cumulative change in which acquired traits are retained is outside of the scope of the theory of natural selection. Even the early evolution of life itself (prior to genetically mediated template replication) cannot be described by natural selection. Specifically, natural selection cannot describe change of state that involves horizontal (Lamarckian) exchange, or occurs through interaction with an incompletely specified context. It cannot describe change wherein variants are evaluated sequentially, and wherein this evaluation can itself change the state space and/or fitness function, because no two variants are ever evaluated according to the same selection criterion. Concerns are also raised as to the methodology used in Simonton's study.

There are few individuals whose contribution to the field of creativity studies is remotely comparable to that of Dean Keith Simonton. His extensive body of work commands and deserves the highest admiration. The single exception to this, in my opinion, is his espousal of a Darwinian view of creativity. In this commentary I show that the flaws in this theory of creativity are so serious that it is not only unlikely, it is impossible.

The goal of Simonton's (2006) paper is to provide experimental evidence for a Darwinian theory of creativity, which he summarizes as follows:

> According to this theory, creativity requires the generation of a certain amount of "blind" ideational variants which are then selected for development into the finished product. These variations are blind in the sense that the creator has no subjective certainty about whether any particular variant represents progress toward the goal rather than retrogression from or diversion away from the goal. As a consequence, the creator must rely on an essentially trial-and-error process that produces more ideas than will ever be used, and will do so in a manner that exhibits no linear or at least no monotonic movement toward the final product.

There are two issues here. First is the question of whether this is an accurate portrayal of the creative process. A second, more fundamental question is: even if this were an accurate portrayal of the creative process, would that imply that the process is Darwinian? It would take a great deal to convince me that the first is true simply because it runs counter to my own experience, and counter to the results of some brilliant experiments on the intuitive antecedents of insight (Bowers et al., 1995). An external manifestation of an unfinished creative idea (such as one of Picasso's sketches for Guernica) is not just the tip of an iceberg, but a ripple made by the tip of an iceberg, and extrapolating from such ripples to the underlying creative enterprise is bound to be dangerous. Undoubtedly as Picasso made externally visible progress on one part of the painting, he was subconsciously busy with another. Thus a lack of continuity in the creative work need not reflect a lack of progressive honing in the internal creative process. But this somewhat 'murky' issue is irrelevant if no matter what the outcome of Simonton's experiment it would not support the theory as claimed. And that I can show that without a doubt is the case. Thus the matters addressed in this commentary are technical in nature.

The commentary begins by demonstrating that Darwin's theory was motivated by a paradox that has no equivalent in creative thought: the paradox of how change accumulates when acquired traits are not inherited. To describe a process of cumulative change in which acquired traits are retained is beyond the scope of the theory of natural selection. Even the early evolution of life itself was not Darwinian, and cannot be described by the theory of natural selection (Gabora, 2006b; Vetsigian et al., 2006). This is followed by a summary of the factors that make natural selection inapplicable to the process by which creative ideas are conceived (elaborated in detail in Gabora, 2005, 2006a) with implications drawn for Simonton's experiment, and his assumption that nonmonotonic variation is indicative of a Darwinian mode of change. Specifically, Finally minor critiques of the experimental methodology are mentioned.

**Darwin was Motivated by a Paradox with No Equivalent in Creative Thought**

What necessitated the theory of natural selection, an intricate theory of *population*-level change, is that acquired traits are not inherited from parent to offspring at the *individual* level in biological lineages. If a cat bites off a rat's tail, it is not the case that the rat's offspring are tail-less. If change keeps getting discarded, how does change accumulate? This is the paradox that biologists faced, and it is this that provided the backdrop against which Darwin's theory solved a pressing dilemma. But there is no such paradox for creativity. Indeed in all domains *other* than biology, explanation of change is straightforward. When an entity undergoes a change of state, say from *p(0)* to *p(1)*, the change is retained. One could say it is 'inherited' by the future states of the entity, *p(2), p(3)* and so forth. The entity does not spontaneously revert back to *p(0)*. For example, if an asteroid collides into a planet, it does not revert back to the state of having not collided into the planet. Similarly, once someone came along with the idea of putting a handle on a cup, cups with handles were here to stay. Since creative ideas *can* retain acquired characteristics, creativity researchers do not share the biologist's need to account for how change occurs despite a discarding of acquired traits each generation. Simonton seeks to borrow a solution to a problem in one field and apply it to another field where that problem does not exist. He wishes to apply natural selection to creativity on the grounds that it is a nonmonotonic process involving trial and error. But it was not this that fueled the theory of natural selection. It was that change (monotonic or not) occurs *despite* the loss of acquired characteristics; *i.e*. despite (one could say) the malfunctioning of the normal mechanism of change. Thus if creative thought exhibits nonmonotonic change, it does not follow that it does so because it is Darwinian. A more parsimonious explanation is that nonmonotonicity can arise not only through natural selection occurring at the population level *but also* through processes occurring at the level of a single entity (such as an unfolding idea) where change is straightforwardly retained.

**Even Biological Evolution Was Not Originally Darwinian**

The problem is not just that when acquired characteristics are retained it is *not necessary* to look to the population level. It is more serious; it is that natural selection is *no longer applicable*. The theory of natural selection theory is not merely vague and descriptive; it has been rendered in precise mathematical terms. For natural selection to be applicable to a process, there must be no inheritance of acquired characteristics (or at least it be negligible compared to change due to differential replication of individuals with heritable variation competing for scarce resources). The periodic 'backtracking' to a previous state observed in biology when one member of a lineage gives birth to another arises because organisms are von Neumann self-replicating automata. Self-replicating automata use a template, a set of instructions encoded in DNA or RNA for how to make a copy of itself. This self-assembly code is both actively transcribed to produce a new individual, and passively copied to ensure that the new individual can itself reproduce. The new individual may change, but the passively copied code within does not.

However creative thoughts do not possess such a code, are not self-replicating automata, and thus retain acquired change. In fact this is also true of early life itself (Gabora, 2006b; Vetsigian et al., 2006). The probability of a self-assembly code such as the genetic code arising spontaneously is exceedingly small; Hoyle infamously compared it to the probability that a tornado blowing through a junkyard would assemble a Boeing 747 (Hoyle, 1981). The implausibility of the spontaneous appearance of a self-assembly code has led to the wide-spread acceptance of *metabolism first* theories, according to which life began with an ensemble of simple, *collectively* replicating molecules, such as an autocatalytically closed[i] set of polymers (Bollobas, 2001; Bollobas & Rasmussen, 1989; Dyson, 1982, 1985; Kauffman, 1993; Morowitz, 1992; Wäechtershäeuser, 1992; Weber, 2000; Williams & Frausto da Silva, 1999, 2002, 2003). Self-replication is not all-at-once (as it is with a self-assembly code), but piecemeal. Although no one molecule replicates itself, the whole is regenerated through the interactions and transformations of its parts. The ensemble is therefore said to be *autopoietic* (Maturana and Varela 1980). Genetically mediated template replication, and thus vertical (as opposed to horizontal) descent emerged subsequently from the dynamics of these molecular systems (Gabora, 2006; Vetsigian et al., 2006). As Vetsigian et al. put it, "the evolutionary process that gave rise to translation [the process by which the genetic code is interpreted to make proteins that make up an organism] is undoubtedly non-Darwinian…. A Darwinian transition corresponds to a state of affaires when sufficient complexity has arisen that the [code] is incapable of tolerating ambiguity, and so there is a distinct change in the nature of the evolutionary dynamics—to vertical descent." As discussed in detail elsewhere (Gabora, 2005) due to the fact that horizontal change generally entails interaction with an incompletely specified context, not only does natural selection on apply to its description, but a different mathematics is needed.

Thus we have (at least) two means by which entities evolve. The one with which we are more familiar is the vertical, highly constrained process of natural selection, which uses a self-assembly code. The one that preceded it, and which has only recently been recognized as a viable means of evolving, is a more haphazard process involving horizontal exchange amongst autopoietic structures. Computer simulations indicate that the non-Darwinian process by which early life evolved exhibited much more nonmonotonic variation than the Darwinian process that followed (Vetsigian et al., 2006). Therefore one cannot conclude that creative thought is Darwinian because it exhibits nonmonotonic variation. Nonmonotonic change is not a litmus test for Darwinism. Indeed there is nothing in natural selection that precludes monotonic change. A genetic algorithm, for example, can exhibit monotonic change.

**Darwinism Cannot Describe Honing**

The Darwinian view of creativity presupposes that an idea is 'stored' in memory waiting to be selected out from amongst a set of others. But as Edelman forcefully points out, human memory works very different from a computer; one does not retrieve an item from memory so much as *reconstruct* it (Edelman, 2000, 2006). Its role in thought is *participatory* (Gabora & Aerts, 2002; Rosch, 1999). An item in memory is never re-experienced in exactly the form it was first experienced. It is colored, however subtly, by what we have experienced in the meantime, re-assembled spontaneously in a way that relates to the task at hand, and if its relevance is unclear it is creatively *redescribed* (Karmiloff-Smith, 1992) from different real or imagined perspectives or contexts until it comes into focus**.** One innovates not by randomly choosing amongst predefined alternatives but by thinking through how something could work. The best a Darwinian approach could hope for is to account for novelty that is both randomly generated and restricted to the same set of properties as what came before, just perhaps to a greater or lesser degree. It cannot account for novelty that is generated strategically, or that makes use of our associative capacities, and it cannot account for the invention of items with *new* properties. As Boden (1990) would put it, they can account for solutions that explore new areas of an existing state space, but not those that break out of a pre-existing conceptual space. Thus it cannot account for the sort of novelty generated by creative minds.

**A Different Fitness Function for Each Variant**

Simonton claims that his critics suffer from the misconception that "the theory only mandates the existence of two or more distinguishable variations that represent alternative directions for future development of an incipient idea (cf. natural selection operating on just two alleles). Hence, a series of sketches of a particular figure only has to contain a minimum of two rival conceptions to be considered part of a Darwinian process." But elsewhere he admits that the ideation process is sequential, one thought at a time:

> Thus, the artist sometimes has false negatives, where he neglects to pursue an idea that will figure prominently in the final painting, and other times he will have false positives where he will temporarily pursue an idea that ultimately gets him nowhere. Whether he goes in the right or wrong direction can be ascribed to chance rather than to intelligence or expertise. Occasionally the artist guesses right while on other occasions he fruitlessly pursues a less lucky hunch. Progress toward the final outcome can thus be described as a sequence of *nonmonotonic variants*.

The fact that these thoughts occur "as a sequence" disqualifies them as candidates for selection theory. This is incompatible with selection theory because it presupposes requires multiple, distinct, simultaneously-actualized states. As Okasha (2001) notes, in attempting to apply selection theory to a temporal sequence, whether something gets selected or not depends arbitrarily on how you break up the sequence. But there is an even deeper problem. Since each thought contributes to the context in which the next is evaluated, never do two thoughts undergo the same 'selective pressure'. For example, the process of generating one sketch affected how Picasso continued from thereon to generate others. Since each successive thought can alter the fitness function itself, they are not evaluated with respect to the same fitness function, and thus there is no basis for selecting amongst them. This yet another reason why creative thought is a process of honing, not selecting.

**Critique of the Methodology**

Even if the hypothesis that creativity is Darwinian were feasible, and even if findings of nonmonotonicity provided more support for this hypothesis than for alternatives, the conclusions of this study would still be questionable because of the methodology used. I was asked by the author to carry out the task of ordering the drawings, but instead of doing it myself I gave it to the students of my Psychology of Creativity class. They did it together, dividing themselves into groups and each group ordering a bundle of sketches, then joining another group and re-ordering the combined bundles, and so forth, until the entire packet of sketches was ordered. The procedure spilled over the class time, in the end taking well over an hour to complete. I was informed by Simonton that the data provided by my class were not used, in part because they arrived after his paper had been submitted, and in part because the class had carried out the task together as a group. These are valid reasons not to include the data. My critique has nothing to do with the integrity of experimenter or the procedure used in an ideal sense but rather with the down to earth fact that even with the best of initial intentions one eventually loses enthusiasm for the task of trying to perfectly order 79 different sketches. After observing students carry out the task of ordering this mountain of drawings, one must ask whether the lack of monotonicity reflects simply a desire to finally be finished with what may appear to be a seemingly endless (not to mention seemingly pointless) task. If the pile had consisted of only 10 or 15 sketches I am certain every effort would have been made to ensure that they were in the best possible order. But with 79, after a certain point one admittedly feels that one has put in sufficient time on the task and must move on to other things, and send the pile as it stands. It must be conceded that as the size of the pile increases, less effort surely goes into ensuring the best possible position for any sketch, and it must also be conceded that the number of sketches that had to be ordered in this experiment would test the patience of all but the most devoted sketch orderers. The fact that only five subjects including the experimenter himself (plus my class) could be convinced to complete the task is telling in this respect.

**Conclusions**

Simonton concludes his paper by claiming "The above results strongly support the contention that the creative process underlying Picasso's *Guernica* was accurately portrayed as Darwinian." This is false. He has provided evidence that the external manifestations of a creative process exhibit a pattern of change that is nonmonotonic. But that does not mean that the creative process is Darwinian, for it violates the conditions that make natural selection applicable to its description. Natural selection was an ingenious solution to a specific problem. The ingenuity of Darwin's approach was to look for a mechanism of change at the level of the population rather than the individual, given that acquired change at the individual level is discarded each generation. It is not applicable to processes where this is not the case. That said, the goal of finding a general framework for understanding and analyzing creative change is worthy. It is hoped that, in Simonton's words, this false positive, this temporarily pursuit an idea that ultimately goes nowhere, is a stepping stone toward more promising theories of how the creative process unfolds.